\documentclass[aps,prl,showkeys,showpacs,reprint,amsmath]{revtex4-1}
\usepackage{amsfonts}
\usepackage{graphicx}
\usepackage{bbm}
\usepackage{bm}
\usepackage{color}
\usepackage[colorlinks]{hyperref}

\hypersetup{urlcolor=blue}
\def\P{\ensuremath{\mathcal{P}}}
\def\id{\ensuremath{\mathbbm{1}}}
\DeclareMathOperator{\Tr}{Tr}
\begin{document}
\title{Correlations between detectors allow violation of the Heisenberg
noise-disturbance principle for position and momentum measurements}
\author{Antonio \surname{Di Lorenzo}}
\affiliation{Instituto de F\'{\i}sica, Universidade Federal de Uberl\^{a}ndia,\\
 38400-902 Uberl\^{a}ndia, Minas Gerais, Brazil}
\pacs{03.65.Ta, 03.67.-a, 05.40.Ca, 06.20.Dk}
\keywords{Uncertainty principle, Noise, Disturbance, Quantum nondemolition measurement}
\begin{abstract}
Heisenberg formulated a noise-disturbance principle stating that there is a tradeoff between noise and disturbance when a measurement of position and a measurement of momentum are performed sequentially, and another principle imposing a limitation on the product of the uncertainties in a joint measurement of position and momentum. We prove that the former, the Heisenberg sequential noise-disturbance principle, holds when the detectors are assumed to be initially uncorrelated from each other, but that it can be violated for some properly correlated initial preparations of the detectors.
\end{abstract}
\maketitle
%
\emph{Introduction.}\textemdash 
The questions that we ask (and answer) in this Letter are of fundamental importance: 
Does the noise-disturbance principle hold, i.e., is it true that increasing the precision in a measurement of position disturbs 
in an uncontrollable way the momentum? How do we quantify the disturbance? What happens if we measure position and momentum not sequentially, but jointly? 

These questions have been investigated theoretically for more than two score years by a minority of physicists, 
but recent experiments \cite{Erhart2012,Rozema2012,Weston2012} have brought them to the attention of the general public. 

Here, we apply Bohr's prescription of only asking questions that can be answered by experiments, however idealized these may be. 
To this end, we consider two nondemolition measurements of position and momentum, and describe both the probes and the 
system quantum-mechanically. Contrary to previous works, we allow the probes to be initially correlated. 

The additional term due to the correlations has surprising consequences: \\
1) It makes possible to violate the Heisenberg uncertainty 
principle, in the noise-disturbance formulation \cite{Heisenberg1927}.
\\
2) 
It voids a recent noise-disturbance relation derived by Ozawa \cite{Ozawa2003}. 
\\
3)
It allows, paradoxically, to counteract the noise introduced in a momentum measurement by preceding it with a position measurement, and vice versa.
\\
The results presented herein are valid for joint or sequential measurements, for any initial preparation of the probes and of the system, and for arbitrary coupling strengths. 

%

Heisenberg \cite{Heisenberg1927} formulated the uncertainty principle in an ambiguous way: either a measurement of position with an uncertainty $\epsilon_X$ entails 
a subsequent disturbance on the momentum $\eta_{P|X}$ such that $\epsilon_X \eta_{P|X}\gtrsim \hbar$, or a joint determination of position and momentum has 
uncertainties $\epsilon_X \epsilon_P\gtrsim \hbar$. 
Furthermore, it is unclear whether Heisenberg was referring to the uncertainty $\epsilon$ on the subensemble of particles for which a given outcome was found, or 
to the uncertainty $\Delta$ over the whole ensemble.  
Soon, Kennard and Weyl \cite{Kennard1927,Weyl1928} proved an inequality for the uncertainties of position and momentum, 
but with a fundamental difference from Heisenberg: 
the Kennard inequality $\sigma_X\sigma_P\ge \hbar/2$ refers to the uncertainties $\sigma_X,\sigma_P$ that one would obtain by measuring $X$ ideally (i.e. with a detector that introduces no noise) on an ensemble of identically prepared particles, and then by measuring $P$ on a distinct ensemble that is prepared in the same way and that has not undergone the $X$ measurement. 


The noise-disturbance principle, the joint uncertainty principle, and the Kennard inequality were confused with one another for a long time. 
A breakthrough  came with a paper by 
Arthurs and Kelly \cite{Arthurs1965}, who considered a joint measurement of position and momentum, assuming a von Neumann protocol  and probes 
prepared in a pure Gaussian state. Arthurs and Kelly found that the quantum nature of the probes increased the total spread of the outputs, 
yielding $\Delta_X \Delta_P \ge \sigma_X \sigma_P + \hbar/2\ge \hbar$. Here, $\sigma$ refers to the intrinsic spread of the system before the measurement, 
and $\Delta$ to the spread of the pointer variables of the probes after the measurement.  We call this relation the Arthurs-Kelly inequality. 
Although with a two-decade delay, the pioneering result of Arthurs and Kelly stimulated further investigations on the uncertainty principle for joint measurements of 
noncommuting variables \cite{Yuen1982,Arthurs1988,Martens1990,Martens1991,Ishikawa1991a,Ishikawa1991b,Raymer1994,Leonhardt1995,Appleby1998,Trifonov2001,DAriano2003,Ozawa2004b,Hall2004,Werner2004,Busch2007}. 

The distinction between Heisenberg noise-disturbance principle and Kennard inequality became gradually clear \cite{Ballentine1970}, 
and it has given rise to a parallel line of research \cite{Ozawa2002,Ozawa2003,Ozawa2003b,Ozawa2004,Busch2004,Vorontsov2005,Lund2010,Distler2012} 
that culminated in establishing a universal noise-disturbance relation, 
\begin{equation}\label{eq:ozawa}
{\epsilon}_X \tilde{\eta}_{P|X} + \epsilon_X \sigma_P + \sigma_X \tilde{\eta}_{P|X} \ge \frac{\hbar}{2}, 
\end{equation}
with $\tilde{\eta}_{P|X}$ representing the disturbance in the $P$ variable, as defined by Ozawa \cite{suppl}.
%
	
\emph{Definitions.}\textemdash
The measuring apparatus consists in two probes, one measuring the position, the other the momentum. 
We introduce the wave number $K=P/\hbar$ and factor out $\hbar$ in the interaction, which is taken to be
\begin{equation}
H_{int}=-\hbar\left[\lambda_X  \delta(t+\tau)  \Hat{\Phi}_X \Hat{X}+\lambda_K  \delta(t-\tau)  \Hat{\Phi}_K \Hat{K}\right].
\label{eq:hint}
\end{equation} 
The interaction is assumed instantaneous, for simplicity, $\Hat{\Phi}_A$  is an operator on the probe $A$, 
and $\tau\in\{+,0,-\}$ specifies the order of the measurements. 
 We labelled each probe with the letter corresponding to the variable that it measures, and let $A\in\{X,K\}$ the generic label. 
The initial state of the probes is assumed to be uncorrelated from that of the system, so that before the interaction 
$\rho=\rho_{pr}\otimes\rho_{sys}$. 
Averages $\langle\dots\rangle$ are meant to be taken with the initial state $\rho$. 
The readout of each probe is described by a family $\Hat{F}_A(\mu)$ of positive operators, with 
$\int d\mu \Hat{F}_A(\mu) = \id$. The readout states are assumed to be classical, in the sense that they commute 
with each other. This defines a preferential basis $|J_A\rangle$ that diagonalizes at once all $\Hat{F}_A(\mu)$. 
Notice that most of the literature considers the case of an ideal readout $\Hat{F}_A(\mu)=|J_A=\mu\rangle\langle J_A=\mu|$. 
In a nondemolition measurement, $\Hat{\Phi}_A$  is the operator of the probe $A$ that 
generates the translations in the preferred basis, i.e., $\exp[ia\Hat{\Phi}_A]|J_A\rangle=|J_A+a\rangle$. 
We indicate by $\Hat{J}_A$ the operators diagonal in the $|J_A\rangle$ basis that satisfy the canonical commutation relation 
$[\Hat{\Phi}_A,\Hat{J}_A]=i$.  
We shall absorb the coupling constants through the canonical scaling $\Phi_A\to\lambda_A \Phi_A$, $J_A\to J_A/\lambda_A$. 
Notice that thus $J_X$, $X$, and $\Phi_K$ have the same dimensions L, while 
$J_K$, $K$, and $\Phi_X$ have dimensions L$^{-1}$.

\begingroup
\squeezetable
\begin{table}[h]
\begin{ruledtabular}
\begin{tabular}{c p{1.9in}}
Symbol&Meaning\\
\hline
$\sigma_A$& Initial uncertainty of $A$\\
$\delta_A$& Initial uncertainty of $J_A$\\
$\tilde{\delta}_A$& Initial uncertainty of $\Phi_A$\\
$\delta'_A$& Resolution of the readout\\
$\Delta_A$& Final uncertainty of $J_A$\\
$\Delta_{A'|A}$& Final uncertainty of $J_{A'}$ when preceded or accompanied by a measurement of $A$\\
$\epsilon_A=\left(\Delta_A^2-\sigma_A^2\right)^{1/2}$& Noise introduced by the probe\\
$\eta_{A'|A}\!=\!\left(\Delta^2_{A'|A}-\Delta^2_{A'}\right)^{1/2}$& Disturbance introduced in a variable $A'$ by a previous measurement of $A$\\
$D_A$& Systematic error in a measurement of $A$\\
$D_{A'|A}$& Systematic disturbance in a measurement of $A'$ introduced by a previous measurement of $A$\\
$\kappa$& Initial covariance between $\Phi_X$ and $J_K$\\
$\xi$& Initial covariance between $\Phi_K$ and $J_X$\\
\end{tabular}
\caption{\label{table} List of symbols for uncertainties and disturbance.}
\end{ruledtabular}
\end{table}
\endgroup

Table \ref{table} resumes the definitions of noise and disturbance given below.
We define the initial intrinsic variances of the system
$\sigma^2_A = \langle \Hat{A}^2\rangle - \langle \Hat{A}\rangle^2$, 
 the initial variances of the probes
$\delta^2_A =\langle \Hat{J}_A^2 \rangle  -\langle \Hat{J}_A\rangle^2$  
and $\tilde{\delta}^2_A = \langle \Hat{\Phi}_A^2 \rangle  -\langle \Hat{\Phi}_A\rangle^2$,
their biases $\langle \Hat{J}_A\rangle$ , 
$\langle\Hat{\Phi}_A\rangle$, 
and their cross-covariances 
\begin{subequations}
\begin{align}
\kappa&=\langle \Hat{\Phi}_X \Hat{J}_K\rangle-\langle \Hat{\Phi}_X \rangle\langle\Hat{J}_K\rangle
,\\
\xi&=\langle \Hat{\Phi}_K \Hat{J}_X\rangle-\langle \Hat{\Phi}_K \rangle\langle\Hat{J}_X\rangle
.
\end{align}
\end{subequations}
Notice that the operators in the covariances commute, so there is no ambiguity.
Recall that Kennard inequality establishes that $\sigma_X \sigma_K\ge 1/2$ and
 $\delta_A\tilde{\delta}_A\ge 1/2$, and that 
the covariances obey the Cauchy--Schwarz inequalities $|\kappa|\le \tilde{\delta}_X\delta_K$ and 
$|\xi|\le \tilde{\delta}_K\delta_X$. 
We also introduce the resolution ${\delta'_A}^2(\mu) =\langle \Hat{J}_A^2\rangle_\mu-\langle \Hat{J}_A\rangle_\mu^2$, 
where $\langle\dots\rangle_\mu$ denotes average in the normalized readout state 
$\rho_{A}(\mu) \equiv\Hat{F}_{A}(\mu)/\Tr[\Hat{F}_{A}(\mu)]$. 

\emph{Noise and disturbance: operational definitions.---}
We wish to discuss the uncertainty relation in the Heisenberg formulation $\epsilon_A \eta_{A'|A}\ge r$, where $r$ is a positive real number 
($r\simeq 1$ in the argumentation of Heisenberg), $\epsilon_A$ is the noise in the variable that is measured first, and 
$\eta_{A'|A}$ is the disturbance introduced in the second measurement by the first one. Thus, we need to agree on the definition of $\epsilon_A$ and $\eta_{A'|A}$. 
For definiteness, say that we measure first $X$, then $K$. 

We shall consider the noise $\epsilon_X$ that is introduced by the measuring apparatus, not that intrinsic to the system. 
We can determine $\epsilon_X$ by calibration: the system is prepared in $\rho_{sys}$ and it is measured by a reference probe that introduces no bias and 
a known noise $\epsilon^{(0)}_X$. 
By measuring the average, we know the intrinsic average $\langle X\rangle_0$ of the system, and by estimating the 
total variance of the outcome $(\Delta^{(0)}_X)^2$, we infer the intrinsic uncertainty of the system $\sigma_X^2=(\Delta^{(0)}_X)^2-(\epsilon^{(0)}_X)^2$. 
Then we repeat the measurement with the probe that we want to characterize on an identically prepared system. 
The readout of the probe gives an average $\langle X\rangle$ and a variance $\Delta_X^2$. 
We define the statistical noise $\epsilon_X^2=\Delta_X^2-\sigma_X^2$ and the systematic error 
$D_X=\langle X\rangle -\langle X\rangle_0$.  
The total error $\mathcal{E}_X=(D_X^2+\epsilon_X^2)^{1/2}$ is the sum in quadrature of the two errors. 
It may be controversial whether the systematic error would be included by Heisenberg in his formulation, or whether 
the noise to consider should be only the statistical one. In any case, it is always possible to make $D_X$ zero while leaving $\epsilon_X$ unchanged. 
For a nondemolition measurement, this operational definition coincides with that given by Ozawa. 

Notice that in statistics, the statistical noise characterizes the precision, and the systematic error the accuracy. 
Furthermore $\epsilon_X$ represents the uncertainty in each individual measurement, that usually is the predictive uncertainty \cite{Appleby1998}, 
while $\Delta_X$ the ensemble uncertainty \cite{Ballentine1970}. 

We define the disturbance in an analogous way as we defined the noise. 
First, we prepare an ensemble of particles in the state $\rho_{sys}$, skip the $X$ measurement ($\lambda_X=0$), 
and measure $K$. We find an average $\langle K\rangle$ with a statistical spread $\Delta_{K}$. 
Then we repeat the procedure, but precede or accompany the $K$-measurement with an $X$-measurement. 
We find an average $\langle K\rangle_{|X}$ with a statistical spread $\Delta_{K|X}$. 
Here, we shall not condition the spread on a specific outcome $X_0$ of the first measurement, even though we could. 
We define the statistical disturbance $\eta^2_{K|X}=\Delta^2_{K|X}-\Delta^2_{K}$. Notice that, a priori, $\eta^2$ may be negative. 
A purely imaginary value for $\eta$ corresponds to the case 
where the ``disturbance'' actually decreases the uncertainty, so that it is not a disturbance at all. 
We also define the systematic disturbance $D_{K|X}=\langle K\rangle_{|X}-\langle K\rangle$.  
Again, it may be controversial whether this second term represents a disturbance as meant by Heisenberg. 
Nonetheless, $D_{K|X}$ can be made zero without changing $\eta_{K|X}$. 
The total disturbance is the sum, in quadrature, of the statistical and systematic disturbances 
$\mathcal{D}_{K|X}=(\eta^2_{K|X}+D^2_{K|X})^{1/2}$. 


\emph{Proof.}\textemdash For definiteness, we consider a measurement of $X$ followed by a measurement of $K$. 
At time $t=-\tau$ a non-demolition measurement of $\Hat{X}$ is made, 
i.e. the system and the probe evolve with the Hamiltonian $H_{-}=-\hbar\delta(t+\tau) \Hat{X}\Hat{\Phi}_X$. 
Thus, on one hand the variable $\Hat{J}_X$ varies according to the Heisenberg equation of motion
\begin{equation}
\frac{d}{dt}\Hat{J}_X=i\delta(t+\tau)\Hat{X} [\Hat{J}_X,\Hat{\Phi}_X] = \delta(t+\tau)\Hat{X},
\end{equation}
so that $\Hat{J}_X(-\tau+0)-\Hat{J}_X(-\tau-0)=\Hat{X}(-\tau)$; on the other hand, the wave number $\Hat{K}$ 
varies as well, giving $\Hat{K}(-\tau+0)-\Hat{K}(-\tau-0)=\Hat{\Phi}_X(-\tau)$. 
Then, at time $t=\tau$ the system interacts with a second probe through 
$H_{+}=-\hbar\delta(t-\tau) \Hat{K}\Hat{\Phi}_K$. This causes a shift in $\Hat{J}_K$: 
\begin{align}
\Hat{J}_K(\tau+0)-\Hat{J}_K(\tau-0)&=\Hat{K}(\tau)\simeq \Hat{K}(-\tau+0)
\nonumber\\
&=\Hat{K}(-\tau-0)+\Hat{\Phi}_X(-\tau)
\end{align}
 for $\tau\to 0^+$. 
Thus, for $\tau\to 0^+$, the final variance in the $J_K$ readout variable is 
\begin{align}
\Delta_{K|X}^2&\equiv\langle \Hat{J}_K(\tau+0)^2\rangle -\langle \Hat{J}_K(\tau+0)\rangle^2
\nonumber
\\
&=\langle \left(\Hat{J}_K(\tau-0)+\Hat{K}(-\tau-0)+\Hat{\Phi}_X(-\tau)\right)^2\rangle
\nonumber 
\\
&\quad -\langle \Hat{J}_K(\tau-0)+\Hat{K}(-\tau-0)+\Hat{\Phi}_X(-\tau)\rangle^2
\nonumber 
\\
&=
\langle\left( \Hat{J}_K+\Hat{\Phi}_X\right)^{\!2}\rangle-\langle\Hat{J}_K+\Hat{\Phi}_X\rangle^2
+\langle\Hat{K}^2\rangle-\langle\Hat{K}\rangle^2
\nonumber 
\\
&=\delta_K^2+\tilde{\delta}_X^2+2\kappa +\sigma_K^2 ,
\label{eq:sequnck0}
\end{align}
where variables with no time argument refer to the instant immediately before the interaction started, 
and we used in the last-but-one line the fact that the system is initially uncorrelated from the probes. 

In a realistic measurement, we should consider that, even if the detector is in the final state $|J\rangle$, 
the output $\mu$ could differ from $J$ with a probability $p(\mu|J)=f(\mu-J)$, where $f(\mu-J)$ is a probability 
centered around 0 and with a variance ${\delta'}^2$. 
Heuristically, we add a further term, due to this finite resolution of the readout state: 
\begin{equation}
\Delta_{K|X}^2=\delta_K^2+\tilde{\delta}_X^2+2\kappa +\sigma_K^2+{\delta_K'}^{2}.
\label{eq:sequnck}
\end{equation}
A rigorous proof, generalizing results presented in Refs.~\cite{DiLorenzo2011,DiLorenzo2013a}, 
that this is the correct formula will be provided elsewhere \cite{DiLorenzo2013e}. 

If no measurement of $X$ was performed, then $\Hat{K}(-\tau+0)-\Hat{K}(-\tau-0)=0$, so that 
\begin{align}
\Delta_{K}^2&\equiv\langle \Hat{J}_K(\tau+0)^2\rangle -\langle \Hat{J}_K(\tau+0)\rangle^2
\nonumber
\\
&=\langle \left(\Hat{J}_K(\tau-0)+\Hat{K}(-\tau-0)\right)^2\rangle
\nonumber 
\\
&\quad -\langle \Hat{J}_K(\tau-0)+\Hat{K}(-\tau-0)\rangle^2
\nonumber 
\\
&=
\langle\Hat{J}_K^{2}\rangle-\langle\Hat{J}_K\rangle^2
+\langle\Hat{K}^2\rangle-\langle\Hat{K}\rangle^2
\nonumber 
\\
&=\delta_K^2+\sigma_K^2 .
\end{align}
We introduce the heuristic correction here as well, yielding
\begin{equation}
\Delta_{K}^2=\delta_K^2+\sigma_K^2+{\delta_K'}^{2}.
\end{equation}
Thus, the statistical disturbance introduced by the $X$ measurement is 
\begin{equation}
\eta_{K|X}^2\equiv \Delta_{K|X}^2-\Delta_K^2=\tilde{\delta}_X^2+2\kappa.
\end{equation}
The statistical variance in the $J_X$ output is simply
\begin{align}
\Delta_{X}^2&\equiv\langle \Hat{J}_X(-\tau+0)^2\rangle -\langle \Hat{J}_X(-\tau+0)\rangle^2
\nonumber
\\
&=\langle \left(\Hat{J}_X(-\tau-0)+\Hat{X}(-\tau)\right)^2\rangle
\nonumber 
\\
&\quad -\langle \Hat{J}_X(-\tau-0)+\Hat{X}(-\tau)\rangle^2
\nonumber 
\\
&=
\langle\Hat{J}_X^{\!2}\rangle-\langle\Hat{J}_X\rangle^2
+\langle\Hat{X}^2\rangle-\langle\Hat{X}\rangle^2
\nonumber 
\\
&=\delta_X^2+\sigma_X^2 ,
\label{eq:sequncx0}
\end{align}
so that, after correcting for the finite resolution, 
\begin{equation}
\Delta_{X}^2=\delta_X^2+\sigma_X^2+{\delta_X'}^{2}.
\label{eq:sequncx}
\end{equation}
The statistical noise is thus 
\begin{equation}
\epsilon_X^2\equiv \Delta_{X}^2-\sigma_X^2=\delta_X^2+{\delta_X'}^{2}.
\end{equation}
The systematic error in $X$ is $D_X=\langle \Hat{J}_X\rangle$ and the systematic disturbance in $K$ is 
$D_{K|X}=\langle \Hat{J}_K\rangle+\langle \Hat{\Phi}_X\rangle-\langle \Hat{J}_K\rangle=\langle \Hat{\Phi}_X\rangle$. 
Clearly, the initial state of the probes can be gauged so that 
$\langle \Hat{J}_X\rangle=\langle \Hat{\Phi}_X\rangle=\langle \Hat{J}_K\rangle=\langle \Hat{\Phi}_K\rangle=0$ without 
affecting variances and covariances. 

Equations~\eqref{eq:sequnck} and \eqref{eq:sequncx} could be derived alternatively by differentiating twice in 
$\chi_K$ and $\chi_X$ the logarithm of Eq. (15) in Ref.~\cite{DiLorenzo2011}. 

\emph{Violation of the noise-disturbance principle.}\textemdash
For initially uncorrelated probes $\kappa=0$, so that $\eta^2_{K|X}=\tilde{\delta}_X^2$. 
The Heisenberg noise-disturbance relation becomes then 
\begin{equation}
\epsilon_X \eta_{K|X} = \tilde{\delta}_X\sqrt{\delta^2_X+\delta'^2_X}  \ge \frac{1}{2}. 
\end{equation}
Thus, if the probes are not correlated, the noise-disturbance relation for position and momentum holds, in agreement with previous results 
\cite{Braginsky1992,Ozawa2002,Ozawa2004}. 

However, if the probes are correlated, 
 $\kappa$ can be negative. The disturbance has the bounds 
\begin{equation}
\tilde{\delta}_X (\tilde{\delta}_X-2\delta_K)\le \eta^2_{K|X}\le \tilde{\delta}_X (\tilde{\delta}_X+2\delta_K). 
\end{equation}
If $\tilde{\delta}_X\le 2\delta_K$, the lower limit is negative, so that the disturbance can be arbitrarily small, 
invalidating Heisenberg's argument. Examples of initial states $\rho_{pr}$ allowing the violation of Heisenberg 
noise-disturbance principle are provided in the Supplemental Material section \cite{suppl}.

Other possibilities leading to the violation of the noise-disturbance relation that were explored in the former literature are the EPR thought-experiment \cite{Einstein1935}, and Ozawa's model \cite{Ozawa2002}. 
Let us discuss them briefly: in the EPR setup, one particle works as a probe, but the two particles are initially correlated. Thus the hypothesis 
that system and probes are initially uncorrelated is violated. Furthermore, due to this limitation, the EPR scheme does not work for any initial state of the system. 
On the other hand, Ozawa violation relies on a single probe whose coordinate $y$ gives information about the momentum $p_x$ of a particle (possibly coinciding with the probe) 
and whose momentum $p_y$ gives information about $x$. 
In this case, one can measure either $y$ or $p_y$, but not both, hence it is questionable whether Ozawa's model produces a genuine violation of the noise-disturbance principle, 
as noise and disturbance refer to different setups.

\emph{Violation of Ozawa relation.}\textemdash
According to Ozawa, Eq.~\eqref{eq:ozawa} holds. 
However, we have demonstrated that $\eta_{K|X}$ can be zero. 
Ozawa's relation then reduces to $\epsilon_X \sigma_K \ge 1/2$. This inequality can be violated, since 
$\epsilon_X$ and $\sigma_K$ are independent of each other, the first referring to the probe, the second to the system. 
This does not show that Ozawa's relation is wrong, as it is mathematically flawless, 
but that the definition of disturbance upon which it relies 
is incompatible with our operational definition, since it disregards the possibility of correlated probes, and therefore it should be discarded \cite{suppl}.

\emph{Joint uncertainty.}\textemdash
Heisenberg also formulated the uncertainty principle in terms of the mutual uncertainty introduced in a joint measurement of  momentum and position.
Thus, let us consider the case of joint measurements, $\tau=0$.
The proof is slightly complicated by the interaction being simultaneous, 
$H_{int}=-\hbar\delta(t)[\Hat{\Phi}_X \Hat{X}+\Hat{\Phi}_K \Hat{K}]$. 
Applying Heisenberg equations of motion, we find that the value of $\Hat{J}_X$ is kicked by an amount $\Hat{X}(0)$. 
However $\Hat{X}$ now has a discontinuity at $t=0$, since the joint measurement of momentum gives 
$\Hat{X}(0+)-\Hat{X}(0-)=-\Hat{\Phi}_K$. An analogous situation happens with the variable $\Hat{J}_K$. 
The correct prescription, which can be proved rigorously \cite{Arthurs1965,DiLorenzo2011}, 
is to consider 
\begin{align}
\Hat{J}_X(0+)-\Hat{J}_X(0-)&=\Hat{X}(0)\equiv\frac{\Hat{X}(0+)+\Hat{X}(0-)}{2}
\nonumber
\\
 &=\Hat{X}(0-)-\frac{1}{2} \Hat{\Phi}_K
\end{align}
 and 
\begin{align}
\Hat{J}_K(0+)-\Hat{J}_K(0-)&=\Hat{K}(0)\equiv\frac{\Hat{K}(0+)+\Hat{K}(0-)}{2} 
\nonumber
\\
 &=\Hat{K}(0-)+\frac{1}{2} \Hat{\Phi}_X.
\end{align}
The variances, including the correction due to the finite resolution, are then 
\begin{align}
\Delta^2_{X} =&\ \sigma^2_X + \delta_X^2 +\delta'^2_X+\frac{1}{4}\tilde{\delta}_K^2-\xi
,
\label{varxjoint}
\\
\Delta^2_{K} =&\ \sigma_K^2 + \delta_K^2 +\delta'^2_K+ \frac{1}{4}\tilde{\delta}_X^2+\kappa 
. 
\label{varkjoint}
\end{align}
Equations~\eqref{varxjoint} and\eqref{varkjoint}, minus the contribution of the finite readout resolution,  
were already derived in Ref.~\cite{DiLorenzo2011}. 
We remark that there is a connection between the noises, due to the Kennard inequality, 
or better, to the canonically invariant generalization proved by Schr\"{o}dinger \cite{Schrodinger1930} and Robertson \cite{Robertson1930}. This is best seen by defining 
\begin{subequations}
\begin{equation}
\Hat{u}_X=\Hat{J}_X-\Hat{\Phi}_K/2-\langle \Hat{J}_X-\Hat{\Phi}_K/2\rangle
\end{equation}
 and 
\begin{equation}
\Hat{u}_K=\Hat{J}_K+\Hat{\Phi}_X/2-\langle\Hat{J}_K+\Hat{\Phi}_X/2\rangle.
\end{equation}  
\end{subequations}  
As $[\Hat{u}_K,\Hat{u}_X]=i$, then, by Kennard inequality, 
$(\delta_X^2+\tilde{\delta}^2_K/4-\xi) (\delta_K^2+\tilde{\delta}^2_X/4+\kappa)\ge 1/4$. 
If we interpret Heisenberg's argument as referring to the noise introduced by the measurement, then 
$\epsilon^2_X\epsilon^2_K =(\delta_X^2 +\delta'^2_X+\frac{1}{4}\tilde{\delta}_K^2-\xi)(\delta_K^2 + \delta'^2_K+\frac{1}{4}\tilde{\delta}_X^2+\kappa)\ge 1/4$. Thus, the joint uncertainty relation holds also when accounting for correlated detectors. 
Arthurs and Kelly, however, interpreted Heisenberg joint uncertainty principle as referring 
to the total uncertainty of the final readout, hence they found \cite{Arthurs1965} a higher limit 
\begin{align}
\Delta_X^2 \Delta_K^2&\ge (\sigma_X^2+\langle \Hat{u}_X^2\rangle) (\sigma_K^2+\langle \Hat{u}_K^2\rangle)
\nonumber
\\
&\ge (\sigma_X^2+\langle \Hat{u}_X^2\rangle) (1/4\sigma_X^2+1/4\langle \Hat{u}_X^2\rangle)\nonumber
\\
&=(2+\langle \Hat{u}_X^2\rangle/\sigma_X^2+\sigma_X^2/\langle \Hat{u}_X^2\rangle)/4\ge 1. 
\end{align}
Notice, that in the EPR thought-experiment \cite{Einstein1935} the measurements can be carried out simultaneously, 
violating the joint uncertainty principle. This is a consequence of the probes and the measured system being initially 
correlated, contrary to what is assumed herein and in Ref.~\cite{Arthurs1965}. 

Furthermore, the mutual disturbances in the joint measurement are $\eta^2_{K|X}=\tilde{\delta}_X^2/4+\kappa$ and 
$\eta^2_{X|K}=\tilde{\delta}_K^2/4-\xi$, which can be made arbitrarily small. 
Hence, while the joint uncertainty principle for the noises $\epsilon_X$ and 
$\epsilon_K$ does hold, the noise-disturbance principle for $\eta_{K|X}$ and $\epsilon_X$, or for $\eta_{X|K}$ and $\epsilon_K$, does not. 

\emph{Cancelling the noise.}\textemdash
Finally, let us go back to the case of sequential measurements, and 
assume perfect anticorrelations  between $\Phi_X$ and $J_K$, so that $\kappa=-\tilde{\delta}_X \delta_K$. 
For instance, we could prepare the probes in an EPR state.
Equation~\eqref{eq:sequnck} becomes then 
\begin{equation}
\Delta^2_{K|X} = \sigma_K^2 + (\delta_K-\tilde{\delta}_X)^2 +\delta'^2_K. 
\label{varkgivx2}
\end{equation}
By making $\delta_K=\tilde{\delta}_X$, the contributions of the probes to the variance partially cancel out, 
leaving only the contribution of the finite resolution, which can be made arbitrarily small, in principle.  
Remember that we concealed the coupling constants by rescaling the variables. Let us restore them momentarily, so that the equality we want to reach is  $\lambda_K^{-1}\delta_K=\lambda_X \tilde{\delta}_X$. This can be realized simply by changing $\lambda_K \lambda_X$, without acting on the initial state of the probes. 
The same considerations hold when the ordering of the measurements is exchanged.

\emph{Conclusions.}\textemdash
We have demonstrated how initial correlations in the detectors invalidate the noise-disturbance principle, and how this can be exploited to reduce the measurement noise.
The key feature, to be discussed at length elsewhere \cite{DiLorenzo2013e}  is that for a given outcome of the first measurement, the correlations are swapped, and pass to be correlations between the measured system and the second probe, 
which allows to counteract the noise introduced by the second measurement, in a sort of negative feedback. 
An interesting question to be investigated is whether classical correlations are sufficient to violate the Heisenberg and Ozawa,  limits, or whether the detectors should be prepared in an entangled state.  
While making a nondemolition measurement of the position of a particle might be a daunting task, an 
equivalent realization with the quadrature components of light may be a viable option in a short-term perspective.
%

\emph{Acknowledgments.}\textemdash
I acknowledge discussions with Jacques Distler. 
This work was performed as part of the Brazilian Instituto Nacional de Ci\^{e}ncia e
Tecnologia para a Informa\c{c}\~{a}o Qu\^{a}ntica (INCT--IQ).


\appendix
\begin{widetext}
\break
\section*{Supplemental Material to\\
Correlations between detectors allow violation of the Heisenberg
noise-disturbance principle for position and momentum measurements}
\section{Preparations of the probes that allow the violation of the noise-disturbance principle}
We consider the probes to be prepared in a pure state $|\psi\rangle$ that has the representation 
\begin{equation}
\langle J_K,\Phi_X|\psi\rangle \propto \exp\left[-\frac{1}{4}(J_K,\Phi_X) C^{-1} (J_K,\Phi_X)^{T}\right],
\end{equation}
where $C$ is the positive-definite covariance matrix 
\begin{equation}
C=
\begin{pmatrix}
\delta_K^2&\kappa\\
\kappa&\tilde{\delta}_X^2.
\end{pmatrix}
\end{equation}
The positive-definiteness is guaranteed by $-\delta_K\tilde{\delta}_X\le\kappa\le \delta_K\tilde{\delta}_X$. 
Hence, letting $\Delta^4=\mathrm{det}(C) = \delta_K^2 \tilde{\delta}_X^2-\kappa^2$, 
\begin{equation}
C^{-1}=\frac{1}{\Delta^4}
\begin{pmatrix}
\tilde{\delta}_X^2&-\kappa\\
-\kappa&\delta_K^2.
\end{pmatrix}
\end{equation}
By construction, we have then $\langle\Phi_X\rangle=\langle J_K\rangle=0$ and 
$\langle\Phi^2_X\rangle=\tilde{\delta}_X^2$, $\langle J_K^2\rangle=\delta_K^2$, $\langle J_K \Phi_X\rangle=\kappa$. 
In the $J_X$ representation, after tracing out the state of the $K$ detector, 
\begin{equation}
\rho_X(J_X,J'_X) \propto \exp\left[-\frac{2\Delta^4}{\delta_K^2}\left(\frac{J_X+J'_X}{2}\right)^2
-\frac{\tilde{\delta}_X^2}{2}\left(J_X-J'_X\right)^2\right].
\end{equation}
Thus $\langle J_X\rangle=0$ and $\langle J_X^2 \rangle ={\delta_K^2}/{4\Delta^4}$. 
For simplicity, we consider the case of an ideal readout $\delta'_X\to 0$. 
Hence the noise in the $X$ measurement is just 
\begin{equation}
\epsilon_X^2=\langle J_X^2 \rangle =\frac{\delta_K^2}{4\Delta^4}. 
\end{equation}

Notice how the Kennard inequality is obeyed, since 
\begin{equation}
\langle J_X^2 \rangle  \langle\Phi^2_X\rangle =\frac{1}{4}\frac{\delta_K^2 \tilde{\delta}_X^2}{\delta_K^2 \tilde{\delta}_X^2-\kappa^2}\ge \frac{1}{4}.
\end{equation}
However, the noise-disturbance relation can be violated. 
The statistical disturbance is indeed 
\begin{equation}
\eta^2_{K|X} = \tilde{\delta}_X^2+2\kappa.
\end{equation}
Let us define the linear correlator $-1\le r\le 1$ through $\kappa= r\delta_K\tilde{\delta}_X$. 
Then the product of noise and disturbance is 
\begin{equation}
\epsilon_X^2\eta^2_{K|X} =\frac{1}{4} \frac{1+2r\tilde{\delta}_X/\delta_K}{1-r^2} .
\end{equation}
It is sufficient to have $\delta_K<2\tilde{\delta}_X$ and $r\le-\delta_K/2\tilde{\delta}_X$ 
in order for the Heisenberg noise-disturbance principle to be violated. 
\section{Comparison between Ozawa's definition of disturbance and the operational definition}
Ozawa considered a system interacting with a probe, which effected a measurement of $\Hat{A}$. 
The disturbance in an observable $\Hat{B}$ was defined by Ozawa as 
\begin{equation}
\tilde{\eta}^2_{B|A}=\Tr\{(U^\dagger\Hat{B}U-\Hat{B})^2\rho_{sys}\otimes\rho_A\} ,
\label{ozawadef}
\end{equation}
with $U$ the time-evolution operator during the interaction between system and probe. 
Thus $U^\dagger\Hat{B}U$ is the time-evolved operator in the Heisenberg picture, after the interaction, 
while $\Hat{B}$ is the operator as it was before the interaction. 
Notice that $U^\dagger\Hat{B}U$ acts on both the Hilbert space of the system and of probe $A$. 
If the operators were directly observable, and if the density matrices represented probability distribution tout court, 
then Eq.~\eqref{ozawadef} would be a natural formula, extending the definition of disturbance of a random variable $x$
due to a process that takes $x$ to $f(x)$: 
\begin{equation}
dist = \int dx [f(x)-x]^2 \Pi(x) .
\end{equation}
However, this is not the case. What is observed are the outputs of the detectors (or better, the modifications of our senses due to these outputs, but there is no need to go further down the von Neumann chain). 
There is a blatant asymmetry in the definition of Eq.~\eqref{ozawadef}: the measurement of $\Hat{A}$ is described, but where is the description of the measurement of $\Hat{B}$? 
Ozawa's definition is implicity assuming that the second probe is making a standard projective measurement, introducing 
no noise. Prima facie, this could seem a desirable property, as a non-projective measurement introduces some 
additional noise. However, as we have demonstrated in the Letter, if the second probe is correlated to the first one, 
the noise can be cancelled. 
Furthermore, let us prove that Eq.~\eqref{ozawadef} does not distinguish between statistical and systematic disturbance. 
Let us consider an alternative measurement that, at the end of the interaction, applies a further transformation $V$ to the system, so that 
\begin{equation}
\tilde{\eta}^2_{B|A}=\Tr\{(U^\dagger V^\dagger\Hat{B}V U-\Hat{B})^2\rho_{sys}\otimes\rho_A\} .
\end{equation}
Let us say that the effect of $V$ is to shift $\Hat{B}$ by a constant amount $b_0$. 
For instance, if $\Hat{B}$ is a position, $V=\exp[i\Hat{P}b_0/\hbar]$. 
Thus $b_0$ is added to the systematic uncertainty. 
Ozawa's definition gives the new disturbance
 \begin{equation}
\tilde{\eta}^2_{B|A}\to \tilde{\eta}^2_{B|A} + b_0^2+2b_0\Tr\{(U^\dagger\Hat{B}U-\Hat{B})\rho_{sys}\otimes\rho_A\}.
\end{equation}
Thus, if the measurement considered earlier does not introduce a systematic disturbance 
$\Tr\{(U^\dagger\Hat{B}U-\Hat{B})\rho_{sys}\otimes\rho_A\}=0$, the new measurement does. In summary, 
Ozawa's definition 
does not distinguish between the systematic and the statistical contribution. 

Finally, let us show that for the non-demolition measurements of position and momentum considered in the Letter, 
our definition of total disturbance reduces to Ozawa's, in the particular case where the probes are assumed to be uncorrelated. 
We have indeed
\begin{equation}
\tilde{\eta}^2_{B|A}=\Tr\{[\exp(-i\Hat{X}\Hat{\Phi}_X)\Hat{K}\exp(i\Hat{X}\Hat{\Phi}_X)-\Hat{K}]^2
\rho_{sys}\otimes\rho_X\} =\langle\Phi_X^2\rangle =\tilde{\delta}_X^2+\langle\Phi_X\rangle^2.
\end{equation}
This is but the total disturbance $\mathcal{D}^2_{K|X}=\tilde{\delta}_X^2+2\kappa+\langle\Phi_X\rangle^2$ 
for $\kappa=0$. 

Our definitions of systematical and statistical disturbance can be written instead
\begin{equation}
D_{K|X}=\Tr[\Hat{J}_K U_{K|X}\rho U_{K|X}^\dagger]- \Tr[\Hat{J}_K U_K\rho U_K^\dagger],
\end{equation}
\begin{equation}
\eta^2_{K|X}=\Tr[\Hat{J}^2_K U_{K|X}\rho U_{K|X}^\dagger]-\Tr[\Hat{J}_K U_{K|X}\rho U_{K|X}^\dagger]^2- 
\left\{\Tr[\Hat{J}^2_K U_K\rho U_K^\dagger]-\Tr[\Hat{J}_K U_K\rho U_K^\dagger]^2\right\},
\end{equation}
where $\rho$ is the density matrix of system and probes, $U_K$ is the time-evolution operator when only the $K$ measurement is performed, and $U_{K|X}$ is the time-evolution operator when the measurement of $K$ is preceded by a measurement of $X$. 

\section{General theory of measurement}
We call measurement any act of inference, even though in a probabilistic sense, about a system through the observation of a second system, the probe, 
that has interacted with the former. 
 What sets apart the probe from the system is its ability to cause a sensation in a human being, if not directly, through a chain of amplifications and further 
interactions with the visible electromagnetic field. Here, the effect of this von Neumann chain \cite{vonNeumann1932} 
is accounted for by considering the readout state of the probe to be a mixed state. 
We shall suppose that the system and the apparatus are initially uncorrelated, so that the initial state is $\Hat{\rho}_{det}\otimes\Hat{\rho}_{sys}$. 
The probability of obtaining an outcome $\mu$ from the probe is given by Born's rule  
\begin{equation}
\P(\mu) =  \Tr\{(\Hat{F}_{det}(\mu)\otimes \id) U (\Hat{\rho}_{det}\otimes\Hat{\rho}_{sys}) U^\dagger \},
\label{eq:probmu}
\end{equation} 
where $U$ is the time-evolution operator, and $\Hat{F}_{det}(\mu)$ 
are a family of positive operators describing the readout states.  Generally, these readout states are mixed, which accounts for having traced out the rest of the von Neumann chain.
They must satisfy the normalization $\int\!d{\mu}\, \Hat{F}_{det}(\mu)=\id$, with $d\mu$ a Lebesgues-Stieltjes measure, corresponding to a discrete or continuous distribution 
of outputs $\mu$. 
The readout states are not normalized, but in general
$\Tr{[\Hat{F}_{det}(\mu)]}=pr(\mu)$. 
The quantities $pr(\mu)$ are the priors \cite{Jaynes2003}, 
and they are not necessarily one. They represent the probability 
of obtaining an outcome $\mu$ in absence of any other information. 
They are not a conventional probability, since 
\begin{equation}
\int\!d\mu\, pr(\mu) =\int d\mu \Tr{[\Hat{F}_{det}(\mu)]} = \Tr{\id} =\infty ,
\end{equation}
unless the probe has a finite-dimensional Hilbert space.
However, the only 
relevant quantity are the ratios $pr(\mu_1)/pr(\mu_2)$ representing relative probabilities. 

Furthermore, it is sensible to assume that the readout states are classical, i.e. $[\Hat{F}_{det}(\mu_1),\Hat{F}_{det}(\mu_2)]=0,\ \forall \mu_1,\mu_2$. 
This hypothesis allows to individuate one, or more, privileged basis, $|J\rangle$, the one that diagonalizes at once all $\Hat{F}_{det}(\mu)$. 
The existence of such a basis can be justified by the decoherence approach \cite{Zurek1981,Zurek1982}. 
The $\Hat{F}_{det}(\mu)$ are labeled by the average of the operator $\Hat{J}=\int\!dJ J\, |J\rangle\langle J|$ 
and they have a spread $\delta'(\mu)$ in $J$, i.e., 
\begin{align}
\mu =&\ \Tr_{det}\{\Hat{J} \Hat{\rho}_{det}(\mu)\},
\label{eq:avmu}\\
\delta'^2(\mu) =&\ \Tr_{det}\{\Hat{J}^2 \Hat{\rho}_{det}(\mu)\} -\left(\Tr_{det}\{\Hat{J}\Hat{\rho}_{det}(\mu)\}\right)^2 ,
\end{align}
with $\Hat{\rho}_{det}(\mu)=\Hat{F}_{det}(\mu)/pr(\mu)$ the normalized readout states.
The spread $\delta'(\mu)$ is but the resolution of the probe, and in principle it can be different for different outputs. 

For instance, we could choose 
\begin{equation}
\Hat{F}_{det}(\mu_n)=\int_{\mu_n-\delta'_{n}/2}^{\mu_n+\delta'_n/2} \!\!\!\!\!dJ \, |J\rangle\langle J|
\end{equation}
and let the readout $\mu$ take discrete values $\mu_n$ spaced by $(\delta'_n+\delta'_{n+1})/2$ from one another 
(then $d\mu$ is a distribution formed by a sum of Dirac deltas). In this case the priors are 
$pr(\mu_n)=\delta'_n$. 

In general, we can write 
\begin{equation}
\Hat{F}_{det}(\mu)=\int\!dJ\, p(\mu|J)\, |J\rangle\langle J|,
\end{equation}
with $p(\mu|J)$ a probability distribution for $\mu$. The priors are thus $pr(\mu)=\int dJ p(\mu|J)$. 
Assuming a uniform resolution $\delta'(\mu)=\delta'$, 
Eq.~\eqref{eq:avmu} is satisfied for $p(\mu|J)=f(\mu-J)$, where $f(\mu)$ is a probability distribution having zero average and spread $\delta'$. An important example is given by the Gaussian-Laplace distribution, 
\begin{equation}
\Hat{F}_{det}(\mu)=\int_{-\infty}^{+\infty} \!\!\!\!\!dJ\, \frac{\exp{[-(J-\mu)^2/2{\delta'}^2]}}{\sqrt{2\pi}\delta'}  |J\rangle\langle J|,
\label{eq:gaussmu}
\end{equation}
with $\mu\in\mathbb{R}$ and uniform priors $pr(\mu)=1$. 

The conditional state of the system for given $\mu$ is 
\begin{equation}
\Hat{\rho}_{sys|\mu} =  \P(\mu)^{-1} \Tr_{det}\{(\Hat{F}_{det}(\mu)\otimes \id) U (\Hat{\rho}_{det}\otimes\Hat{\rho}_{sys}) U^\dagger\}.
\label{eq:condmu}
\end{equation}
Eequivalent formulas arise when treating postselected weak measurement \cite{Aharonov1988,DiLorenzo2012a}, 
but with the difference that $U$ is expanded in a perturbation series and the r\^{o}les  of system and probe are reversed.  
Let us introduce the preparation basis $|I\rangle$, the one in which $\Hat{\rho}_{det}=\int \!dI\, w(I) |I\rangle\langle I|$. 
We can rewrite the conditional state of the system
\begin{equation}
\Hat{\rho}_{sys|\mu} =  \frac{\int dJ dI M_{J,I}(\mu) \Hat{\rho}_{sys}M_{J,I}^\dagger(\mu)}
{\int dJ dI\Tr_{sys}\{E_{J,I}(\mu) \Hat{\rho}_{sys}\}} .
\label{eq:probmu2}
\end{equation} 
We defined the generalized operations
\begin{equation}
M_{J,I}(\mu) = \sqrt{p(\mu|J) w(I)} \langle J|\Hat{U}|I\rangle ,
\label{eq:op}
\end{equation}
and the generalized effects $E_{J,I}(\mu)=M^\dagger_{J,I}(\mu)M_{J,I}(\mu)$. 
Both $M_{J,I}(\mu)$ and $E_{J,I}(\mu)$ are operators on the Hilbert space of the system alone. 
If we do not make the hypothesis of classical readout, Eq. \eqref{eq:op} becomes 
$M_{J,I}(\mu) = \sqrt{p(\mu|J) w(I)} \langle \mu:J|\Hat{U}|I\rangle$, where $|\mu:J\rangle$ is the basis that diagonalizes $\Hat{F}_{det}(\mu)$. 

The measurement can also be described in the language of positive-operator valued measures (POVM) \cite{Davies1976,Helstrom1976,Holevo1982,Kraus1983,Ludwig1983,Busch1995}, even though I am not fond of this 
axiomatic approach that conceals most of the physics behind the formalism. 
The linear operations 
\begin{equation}
\mathcal{I}_{D}(\rho_{sys})\equiv \int_D d\mu\int dJ dI M_{J,I}(\mu) \Hat{\rho}_{sys}M_{J,I}^\dagger(\mu), 
\label{eq:instr}
\end{equation}
where $D$ is a measurable domain in the space of possible outcomes 
(i.e. an element of a $\sigma$--algebra in the Kolmogorov formalism), are called \emph{instruments} if they preserve 
the non-negativity of $\rho$.
In the POVM formalism, the instruments are introduced as the basic objects describing the measurement. 
Besides preserving the nonnegativity of the density operator, they must also satisfy \cite{Davies1970,Davies1976}
\begin{enumerate}
\item
$\mathcal{I}_\emptyset=0$, i.e., the measurement always gives an outcome (notice, however, that the outcome is not necessarily a number, but it could consist, e.g., in the observation that the apparatus has failed, etc.).
\item $\mathcal{I}_{\cup D_j}=\sum_j \mathcal{I}_{D_j}$ for any family of disjoint elements of the $\sigma$--algebra. 
\item $\Tr_{sys}[\mathcal{I}_\Omega(\rho_{sys})]=1$, where $\Omega$ is the whole set of possible outcomes.
\end{enumerate}
The instruments defined in Eq.~\eqref{eq:instr} automatically satisfy the three above properties. 
An interesting question is whether any instrument can be written as in Eq.~\eqref{eq:instr}. 

The instruments describe the change of the density matrix one assigns to the system, conditioned 
on the information that the outcome of the measurement was in $D$. 
In other words, the new density matrix is 
\begin{equation}
\rho_{sys|D}=\frac{\mathcal{I}_\mathcal{D}(\rho_{sys})}{P(D|\rho_{sys})}, 
\end{equation}
with the normalization $P(D|\rho_{sys})$ being the probability of obtaining an outcome in $D$, 
\begin{align}
P(D|\rho_{sys}) =& \Tr_{sys}[E(D)\rho_{sys}],\\
E(D)=&\mathcal{I}_D^*(\id) .
\end{align}
The family of conjugate operators $\mathcal{I}_D^*$is defined univocally by 
\begin{equation}
\Tr_{sys}[\rho \mathcal{I}_D^*(\sigma)] = \Tr_{sys}[\mathcal{I}_D(\rho)\sigma].
\end{equation}

When both $\Hat{F}_{det}(\mu)$ and $\Hat{\rho}_{det}$ are pure states, 
then only one term survives in the double sum of Eq.~\eqref{eq:instr} and the instruments are called L\"{u}ders instruments.
The operators $E(\mu)$ can be inferred from the observed probabilities $\P(\mu)$ by changing the preparation  $\Hat{\rho}_{sys}$. 
Recovering $E_{J,I}(\mu)$, however, is a nontrivial task, and requires the knowledge of the measurement apparatus, 
as different $M_{J,I}(\mu)$ correspond to the same instrument. 
If $E_{J,I}(\mu)$ are known, the operations are determined modulo a unitary operator: 
$M_{J,I}(\mu) =V_{J,I}(\mu) E_{J,I}^{1/2}(\mu)$, where $E^{1/2}_{J,I}$ is univocally defined as a positive operator. 
In other words, we consider the eigenvalues of the positive operator $E_{J,I}(\mu)$, take their arithmetic square root, 
and define $E_{J,I}^{1/2}(\mu)$ as the operators having the 
same eigenstates as $E_{J,I}(\mu)$, and 
as eigenvalues the said arithmetic root. 
The additional operations $V_{J,I}(\mu)$ represent an unwanted and avoidable feedback on the system.
For instance, in a Stern-Gerlach apparatus, the beam exiting the magnetic field gradient could pass through a uniform magnetic field, which rotates the spin, so that a second measurement would not confirm the result of the first one.
Ideally, $V_{J,I}(\mu)=\id$. 
Only by making subsequent measurements on 
the system, is it possible to check whether the measurement process is introducing this unwanted feedback. 

We are interested in particular to the case of two sequential measurements. 
A probe measuring $\Hat{A}$ couples first to the system, then, after the first probe decouples, a second probe 
measuring $\Hat{B}$ couples. 
The time-evolution operator factorizes then as 
$U=U_B U_A$, where $U_A$ (respectively, $U_B$) acts on the Hilbert space of the system 
and the probe $A$ (respectively, $B$). 
As we are making two measurements, it is natural to assume that the readout states factor, 
$\Hat{F}_{det}(\mu)=\Hat{F}_{A}(\mu_A) \Hat{F}_{B}(\mu_B)$, each factor acting on a different Hilbert space. 
A fundamental consideration is that, if we assume as well that the initial state of the two probes factors as 
$\rho_{det}=\rho_A\otimes \rho_B$, i.e. if we assume no initial correlations between the probes, then 
the probability of observing the outcomes $\mu_A,\mu_B$ can be written 
\begin{equation}
\P(\mu_A,\mu_B) = 
\Tr_{sys}\left\{\mathcal{I}_{\mu_B}\left[ 
\mathcal{I}_{\mu_A}(\rho_{sys})\right]\right\}
\end{equation}
and the conditional state of the system 
is 
\begin{equation}
\rho_{sys|\mu}\propto\mathcal{I}^{B}_{\mu_B}\left[\mathcal{I}^{A}_{\mu_A}(\rho_{sys})\right],
\end{equation}
where the composing instruments are defined by 
\begin{align}
\mathcal{I}^{k}_{\mu_k}(\rho)=& \int dI_k dJ_k M^k_{J_k,I_k}(\mu_k) \rho  M^{k\dagger}_{J_k,I_k}(\mu_k) ,
\label{eq:nocorrseq}
\\
M^k_{J,I}(\mu) =& \sqrt{p(\mu|J) w(I)} \langle J|\Hat{U}_k|I\rangle 
\label{eq:op2}
.
\end{align}
However, if we do assume initial correlations, then the total instrument cannot be written as a composition of 
two instruments, contrary to what is commonly assumed \cite{Carmeli2011}, and we can only associate one set 
of instrument, defined on the probability space $M_A\times M_B$.  

Finally, we note that the marginal probabilities are 
\begin{align}
\P(\mu_A) &= \int d\mu_B\Tr\left\{\left[\id\otimes\Hat{F}_{A}(\mu_A) \otimes\Hat{F}_{B}(\mu_B)\right] U_B U_A (\rho_{sys}\otimes\rho_{\mathrm{det}}) U_A^\dagger U_B^\dagger\right\} 
\nonumber
\\
&=
\Tr_{sys,A}\{[\id\otimes F_A(\mu_A)] U_A(\rho_{sys}\otimes\rho_A) U^\dagger_A\} \ , \qquad  \text{with }  \rho_A=\Tr_B(\rho_{det})\\
\P(\mu_B) &= \int d\mu_A\Tr\left\{\left[\id\otimes\Hat{F}_{A}(\mu_A) \otimes\Hat{F}_{B}(\mu_B)\right] U_B U_A (\rho_{sys}\otimes\rho_{\mathrm{det}}) U_A^\dagger U_B^\dagger\right\} 
\nonumber \\
&=
\Tr_{sys,B}\{[\id\otimes\Hat{F}_B(\mu_B)] U_B \rho_{sys,B}U_B^\dagger\} \ , \qquad \text{with } \rho_{sys,B}=\Tr_A [U_A(\rho_{sys}\otimes\rho_{det})U_A^\dagger].
\end{align}
Notice that the correlations are swapped: from correlations between $A$ and $B$, they become correlations between $B$ and the system, after tracing out the first probe. 
Applying Bayes' rule, we see that neither conditional probability 
\begin{align}
\P(\mu_B|\mu_A) &= \frac{\P(\mu_A,\mu_B)}{\P(\mu_A)}=
\frac{\Tr[E(\mu_A,\mu_B) \rho_{sys}]}{\Tr_{sys,A}\{[\id\otimes F_A(\mu_A)] U_A(\rho_{sys}\otimes\rho_A) U^\dagger_A\}} \ , \\
\P(\mu_A|\mu_B) &= \frac{\P(\mu_A,\mu_B)}{\P(\mu_B)}=
\frac{\Tr[E(\mu_A,\mu_B) \rho_{sys}]}{\Tr_{sys,B}\{[\id\otimes\Hat{F}_B(\mu_B)] U_B \rho_{sys+B}U_B^\dagger\}} \ ,
\end{align}
admits a conditional state $\rho_{sys|\mu_A}$ or $\rho_{sys|\mu_B}$ such that 
\begin{align}
\P(\mu_B|\mu_A) &=
\Tr_{sys,B}\{[\id\otimes\Hat{F}_B(\mu_B)] U_B(\rho_{sys|\mu_A}\otimes\rho_B)U_B^\dagger\} \ , \\
\P(\mu_A|\mu_B) &= 
\Tr_{sys,A}\{[\id\otimes\Hat{F}_A(\mu_A)] U_A(\rho_{sys|\mu_B}\otimes\rho_A)U_A^\dagger\} \ , \\
\ ,
\end{align}
unless $\rho_{det}=\rho_A\otimes\rho_B$. 
This is because, for a given output $\mu_A$ (respectively, $\mu_B$), 
the system becomes correlated with the probe $B$ (respectively, $A$). 
\end{widetext}

%

\end{document}